\documentclass[a4paper,11pt]{article}\usepackage{jcappub}\def\Styleopt#1#2{#1}
\usepackage{cmap}
\usepackage[utf8]{inputenc}
\usepackage[T1]{fontenc}

\def\imo{i}
\def\re#1{Re(#1)}
\def\im#1{Im(#1)}
\def\K{{\cal K}}
\def\Order#1{{\cal O}\left(#1\right)}

\begin{document}

\title{Correspondence between grey-body factors and quasinormal modes}

\Styleopt{%
\author[\dagger]{R. A. Konoplya,}
\affiliation[\dagger]{Research Centre for Theoretical Physics and Astrophysics, \\ Institute of Physics, Silesian University in Opava, \\ Bezručovo náměstí 13, CZ-74601 Opava, Czech Republic}
\emailAdd{roman.konoplya@gmail.com}
\author[\ddagger]{A. Zhidenko}
\affiliation[\ddagger]{Centro de Matemática, Computação e Cognição (CMCC),\\ Universidade Federal do ABC (UFABC), \\ Rua Abolição, CEP: 09210-180, Santo André, SP, Brazil}
\emailAdd{olexandr.zhydenko@ufabc.edu.br}
\arxivnumber{2406.11694}
}{%
\author{R A Konoplya$^\dag$ and A Zhidenko$^\ddag$}
\address{$^\dag$ Research Centre for Theoretical Physics and Astrophysics, \\ Institute of Physics, Silesian University in Opava, \\ Bezručovo náměstí 13, CZ-74601 Opava, Czech Republic}
\ead{roman.konoplya@gmail.com}
\address{$^\ddag$ Centro de Matemática, Computação e Cognição (CMCC),\\ Universidade Federal do ABC (UFABC), \\ Rua Abolição, CEP: 09210-180, Santo André, SP, Brazil}
\ead{olexandr.zhydenko@ufabc.edu.br}
}%

\abstract{
Quasinormal modes and grey-body factors are spectral characteristics corresponding to different boundary conditions: the former imply purely outgoing waves to the event horizon and infinity, while the latter allow for an incoming wave from the horizon, thus describing a scattering problem. Nevertheless, we show that there is a link between these two characteristics. We establish an approximate correspondence between the quasinormal modes and grey-body factors, which becomes exact in the high-frequency (eikonal) regime. We show that, in the eikonal regime, the grey-body factors of spherically symmetric black holes can be remarkably simply expressed via the fundamental quasinormal mode, while at smaller $\ell$, the correction terms include values of the overtones. This might be interesting in the context of the recently observed connection between grey-body factors and the amplitudes of gravitational waves from black holes.
The correspondence might explain why grey-body factors are more stable, i.e. less sensitive, than higher overtones to small deformation of the effective potential.
}

\maketitle

\section{Introduction}

The two spectral problems are frequently studied when discussing radiation processes around black holes: the {\it quasinormal mode}  spectrum \cite{Konoplya:2011qq,Kokkotas:1999bd} and the classical scattering problem governed by the {\it grey-body factors}.
In \cite{Oshita:2023cjz} it was noticed that the black hole greybody factor, $\Gamma_{\ell m} (\omega)$, could be an important characteristic for the  estimation of the remnant parameters from gravitational-wave ringdown. There it was shown that for $(\ell,m)=(2,2)$, $\Gamma_{\ell m}$ can be imprinted on the gravitational-wave spectral amplitude $|\tilde{h}_{\ell m} (\omega)|$ in $\omega \gtrsim f_{\ell m} \equiv \re{\omega_{\ell m0}}$ with the form of
\begin{equation}
|\tilde{h}_{\ell m} (\omega)| \simeq c_{\ell m} \times \gamma_{\ell m} (\omega) \equiv c_{\ell m} \times \sqrt{1-\Gamma_{\ell m} (\omega)}/\omega^3 \ \mbox{for} \ \omega \gtrsim f_{\ell m},
\label{main}
\end{equation}
where $\omega$ is a frequency of a gravitational wave and $c_{\ell m}$ is a constant corresponding to the gravitational-wave amplitude.
Furthermore, it was observed that, unlike the first few overtones of the quasinormal spectra, grey-body factors are stable; i.e., they deviate slightly under relatively small deformations of the Kerr geometry \cite{Rosato:2024arw,Oshita:2024fzf}.
Therefore, grey-body factors may be valuable characteristics not only for estimating the intensity of Hawking radiation but also in the context of astrophysical observations of gravitational waves from black holes.

The grey-body factors and quasinormal modes are quantities characterizing problems with different boundary conditions. Therefore, the connection between them is rather surprising. Here, we considered both quantities using the WKB method, assuming the black hole is spherically symmetric and asymptotically flat or de Sitter. It turns out that there is a simple correspondence between the grey-body factors and the fundamental quasinormal mode in the high-frequency (eikonal) regime. In this regime, the WKB method is exact, and therefore the correspondence holds exactly, at least for ``well-behaved'' effective potentials that have a single maximum and monotonically decay towards the event horizon and infinity. The WKB series converges only asymptotically, so extending the correspondence beyond the eikonal regime does not strictly guarantee better accuracy. Nevertheless, including higher-order correspondence, which involves the first few overtones, leads to a considerable improvement in accuracy for lower multipoles $\ell$.

Our work is organized as follows. In Sec.~\ref{sec:WKB}, we review the WKB approach. Sec.~\ref{sec:eikonal} is devoted to deriving the correspondence between grey-body factors and quasinormal modes in the eikonal regime. In Sec.~\ref{sec:correction}, we extend this correspondence beyond the eikonal regime and illustrate it using examples of gravitational perturbations of the Schwarzschild black hole in Sec.~\ref{sec:results}. Finally, in Sec.~\ref{sec:conclusions}, we summarize the obtained results and discuss open questions.

\section{WKB approach}\label{sec:WKB}

When the effective potential $V(r_*)$ in the wavelike equation,
\begin{equation}\label{wavelike}
\frac{d^2\Psi}{dr_*^2}+(\omega^2-V(r_*))\Psi(r_*)=0,
\end{equation}
has a form of a barrier with a single peak, the WKB formula is appropriate for obtaining the dominant quasinormal modes, satisfying the boundary conditions,
\begin{equation}\label{boundaryconditions}
\Psi(r_*\to\pm\infty)\propto e^{\pm\imo \omega r_*},
\end{equation}
which are purely ingoing wave at the horizon ($r_*\to-\infty$) and purely outgoing wave at spatial infinity ($r_*\to\infty$) or cosmological horizon.

In this paper we assume that all the modes are decaying, so that $\im{\omega}<0$, and also consider only positive values of the real oscillation frequencies $\re{\omega}>0$.

The WKB method relies on matching the asymptotic solutions, which satisfy the quasinormal boundary conditions (\ref{boundaryconditions}), with the Taylor expansion around the peak of the potential barrier through the two turning points, which satisfy the equation
$$V(r_*)=\omega^2.$$ The first-order WKB formula corresponds to the eikonal approximation and becomes exact in the limit $\ell\to\infty$. The general WKB expression for the quasinormal frequencies can then be written as an expansion around the eikonal limit as follows \cite{Konoplya:2019hlu}:
\begin{eqnarray}\label{WKBformula-spherical}
\omega^2&=&V_0+A_2(\K^2)+A_4(\K^2)+A_6(\K^2)+\ldots\\\nonumber
&-&\imo\K\sqrt{-2V_2}\left(1+A_3(\K^2)+A_5(\K^2)+A_7(\K^2)\ldots\right).
\end{eqnarray}

The matching conditions for the quasinormal modes require that
\begin{equation}
\K=n+\frac{1}{2}, \quad n=0,1,2,\ldots,
\end{equation}
where $n$ is the overtone number, $V_0$ is the value of the effective potential at its maximum, $V_2$ is the value of the second derivative of the potential at this point, and $A_i$ for $i=2, 3, 4, \ldots$ are the $i$-th order WKB correction terms beyond the eikonal approximation, depending on $\K$ and the derivatives of the potential at its maximum up to the order $2i$. The explicit forms of $A_i$ can be found in \cite{Iyer:1986np} for the second and third WKB orders, in \cite{Konoplya:2003ii} for the 4th-6th orders, and in \cite{Matyjasek:2017psv} for the 7th-13th orders.

For the scattering problem, we consider the same wave equation (\ref{wavelike}),
\begin{equation}\label{wavelike-scatter}
\frac{d^2\Psi}{dr_*^2}+(\Omega^2-V(r_*))\Psi(r_*)=0,
\end{equation}
under boundary conditions that permit incoming waves from infinity. Due to the symmetry of scattering properties, this is equivalent to the scattering of a wave originating from the horizon. Therefore, the scattering boundary conditions are:
\begin{equation}
\begin{array}{rclcl}
\Psi &=& e^{-i\Omega r_*} + R e^{i\Omega r_*}, &\quad& r_*\to+\infty, \\
\Psi &=& T e^{-i\Omega r_*}, &\quad& r_*\to-\infty,
\end{array}
\end{equation}
where $R$ and $T$ are the reflection and transmission coefficients.
From this point onward, we should distinguish the real frequency $\Omega$ in the scattering problem from the complex quasinormal frequency, which is denoted by the lowercase letter $\omega$. The latter usually has an index, $\omega_n$, indicating the overtone number.

When solving the scattering problem with the help of the WKB method, the matching conditions allow us to express the reflection and transmission coefficients as follows \cite{Iyer:1986np},
\begin{eqnarray}\label{reflection}
|R|^2&=&\frac{1}{1+e^{-2\pi\imo \K}},\qquad 0<|R|^2<1.\\
\label{transmission}
|T|^2&=&\frac{1}{1+e^{2\pi\imo \K}}=1-|R|^2.
\end{eqnarray}
where $\K$ is a function of the frequency $\Omega$, defined through the relation given by the WKB formula~(\ref{WKBformula-spherical}).

In order to obtain the reflection/transmission coefficients, equation~(\ref{WKBformula-spherical}) is usually solved numerically for a given $\Omega$. Here we develop an analytic approach for calculation of $\K$ as a function of real $\Omega$ based on the values of the dominant quasinormal frequencies.

\section{Eikonal approximation}\label{sec:eikonal}

For test fields and gravitational perturbations in a spherically symmetric background, the effective potential can be expressed as follows:
\begin{equation}\label{potential-multipole}
V(r_*)=\ell^2U_0(r_*)+\ell U_1(r_*)+U_2(r_*)+\ell^{-1}U_3(r_*)+\ldots,
\end{equation}
where $\ell$ is the multipole number, with its minimum value equal to the spin of the field under consideration $s$.

The eikonal approximation can be obtained from the first-order WKB formula \cite{Schutz:1985km},
$$\Omega^2=V_0-\imo\K\sqrt{-2V_2}+\Order{\ell^{-1}},$$
which, after substituting (\ref{potential-multipole}), reads
\begin{eqnarray}\label{WKBformula-eikonal}
\Omega&=&\sqrt{V_0-\imo \K\sqrt{-2V_2}}+\Order{\ell^{-1}}=\ell\sqrt{U_{00}}-\imo \K\sqrt{\frac{-U_{02}}{2U_{00}}}+\Order{\ell^{-1}},
\end{eqnarray}
where $U_{00}$ is the value of the function $U_0(r_*)$ in its maximum, $U_{02}$ is the value of its second derivative in this point.

This formula usually provides a good approximation for the dominant quasinormal modes, i.e. for $n=0$ or $\K=1/2$,
\begin{eqnarray}\label{WKBformula-eikonal-dominant}
\omega_0&=&\ell\sqrt{U_{00}} - \frac{\imo}{2}\sqrt{\frac{-U_{02}}{2U_{00}}}+\Order{\ell^{-1}},
\end{eqnarray}
so that
$$
\begin{array}{rcl}
\re{\omega_0}&=&\ell\sqrt{U_{00}}+\Order{\ell^{-1}},\\
\im{\omega_0}&=&-\dfrac{1}{2}\sqrt{\dfrac{-U_{02}}{2U_{00}}}+\Order{\ell^{-1}}.\\
\end{array}
$$

At the same time, from (\ref{WKBformula-eikonal}), one can express the value of $\K$ as a function of the real frequency $\Omega$ through the values of the real and imaginary part of the fundamental quasinormal mode $\omega_0$, which is given by eq.~(\ref{WKBformula-eikonal-dominant}),
\begin{eqnarray}\label{eikonal-K}
-\imo\K&=&\frac{\Omega^2-\ell^2U_{00}}{\ell\sqrt{-2U_{02}}} + \Order{\ell^{-1}} = -\frac{\Omega^2-\re{\omega_0}^2}{4\re{\omega_0}\im{\omega_0}} +\Order{\ell^{-1}},
\end{eqnarray}
from which we derive the expression for the transmission coefficient (\ref{transmission}),
\begin{eqnarray}\label{transmission-eikonal}
\Gamma_{\ell}(\Omega)\equiv|T|^2&=&\left(1+e^{2\pi\dfrac{\Omega^2-\re{\omega_0}^2}{4\re{\omega_0}\im{\omega_0}}}\right)^{-1} + \Order{\ell^{-1}}.
\end{eqnarray}
This expression, connecting the grey-body factors $\Gamma_{\ell}(\Omega)$ with the fundamental quasinormal mode $\omega_0$ of a spherically symmetric black hole, is exact in the eikonal limit $\ell \to \infty$. However, for small multipole numbers $\ell$, eq.~(\ref{transmission-eikonal}) is approximate. We further demonstrate that this correspondence can be extended beyond the eikonal regime by incorporating corrections from the overtones.

\section{Higher-order correction to the eikonal formula}\label{sec:correction}

Following \cite{Konoplya:2023moy}, we use the second-order WKB formula applied to the effective potential (\ref{potential-multipole}) and obtain the correction to the expression (\ref{eikonal-K}),
\begin{eqnarray}\label{beyond-eikonal-K-1}
-\imo\K&=&-\frac{\Omega^2-\re{\omega_0}^2}{4\re{\omega_0}\im{\omega_0}}+\Delta_1,
\end{eqnarray}
where $\Delta_1=\Order{\ell^{-1}}$ depends on the value $U_1(r_*)$ in the point of maximum of $U_0(r_*)$ and higher derivatives of $U_0(r_*)$. However, in terms of the values of the dominant quasinormal mode $\omega_0$ and the first overtone $\omega_1$, the expression takes a particularly simple form:
\begin{eqnarray}
\Delta_1&=&\frac{\re{\omega_0}-\re{\omega_1}}{16\im{\omega_0}}+\Order{\ell^{-2}},
\end{eqnarray}
where
$\omega_1$ is the first overtone. Note that we use the second-order formula WKB to obtain the approximations for the values of $\omega_0$ and $\omega_1$ and neglect the terms of the order $\Order{\ell^{-2}}$.

Similarly, with the help of the third-order WKB formula, we obtain the second-order correction,
\begin{eqnarray}\label{beyond-eikonal-K-2}
-\imo\K&=&-\frac{\Omega^2-\re{\omega_0}^2}{4\re{\omega_0}\im{\omega_0}}+\Delta_1+\Delta_2,
\end{eqnarray}
where
\begin{eqnarray}\hspace{-20pt}
\Delta_2&=&-\frac{\Omega^2-\re{\omega_0}^2}{32\re{\omega_0}\im{\omega_0}}\left(\frac{(\re{\omega_0}-\re{\omega_1})^2}{4\im{\omega_0}^2}-\frac{3\im{\omega_0}-\im{\omega_1}}{3\im{\omega_0}}\right)
\\\nonumber&&+\frac{(\Omega^2-\re{\omega_0}^2)^2}{16\re{\omega_0}^3\im{\omega_0}}\left(1+\frac{\re{\omega_0}(\re{\omega_0}-\re{\omega_1})}{4\im{\omega_0}^2}\right)+\Order{\ell^{-3}}.
\label{Delta2}
\end{eqnarray}

The second term in the above expression (\ref{Delta2}) is of the same order of $\ell$ as $\Delta_1$, despite it is obtained from the higher-order WKB formula. It can be easily explained if we recall that one has $\re{\omega_0}=\Order{\ell}$ and the following relation is fulfilled by assumption $\Omega^2-\re{\omega_0}^2=\Order{\ell}$ (see Sec. 4.1 in \cite{Konoplya:2023moy} for a detailed discussion). Notice, that the expression truncated at the the second order has wrong asymptotic behavior as $\Omega\to\infty$. Therefore, the second-order formula, providing a better approximation for small and moderate values of $\Omega$ gives wrong results for $\Omega\gg\re{\omega_0}$. Fortunately, in that regime the grey-body factors are practically indistinguishable from unity.

In order to amend the final formula we employed the six-order WKB formula to obtain the correction proportional to $(\Omega^2-\re{\omega_0}^2)^3$, which is of the order $\Order{\ell^{-2}}$,
\begin{eqnarray}\hspace{-20pt}
\Delta_f&=&-\frac{(\Omega^2-\re{\omega_0}^2)^3}{32\re{\omega_0}^5\im{\omega_0}}\Biggl(1+\frac{\re{\omega_0}(\re{\omega_0}-\re{\omega_1})}{4\im{\omega_0}^2}
\\\nonumber&&
+\re{\omega_0}^2\left(\frac{(\re{\omega_0}-\re{\omega_1})^2}{16\im{\omega_0}^4}-\frac{3\im{\omega_0}-\im{\omega_1}}{12\im{\omega_0}}\right)\Biggr)+\Order{\ell^{-3}}.
\label{Deltaf}
\end{eqnarray}

Although we did not check the higher than sixth order WKB corrections, we believe that the resulting formula contains all the terms of the order $\Order{\ell^{-2}}$,
\begin{eqnarray}\label{beyond-eikonal-K-f}
-\imo\K &=& -\frac{\Omega^2-\re{\omega_0}^2}{4\re{\omega_0}\im{\omega_0}}+\Delta_1+\Delta_2+\Delta_f+\Order{\ell^{-3}}.
\end{eqnarray}
The amended formula resulting from the WKB expansion at the sixth order restores the correct asymptotic behavior for $\Omega\gg\re{\omega_0}$.

The relation (\ref{beyond-eikonal-K-f})\footnote{We share the resultant formula in the ancillary Mathematica\textregistered{} Notebook, which is available from \url{https://arxiv.org/src/2406.11694/anc}.} is the principal result of the present paper. It allows one to get quite a good approximation for the grey-body factors for large values of the multipole number $\ell$. It is important to understand that, although we have used the expansion for large values of the multipole number, the final formula does not explicitly depends on $\ell$, since all the dependence on the particular behavior of the effective potential is given through the values of the two dominant quasinormal frequencies, $\omega_0$ and $\omega_1$. It means that the parameter $\ell$ in eq.~(\ref{beyond-eikonal-K-f}) is merely a bookkeeping parameter of the steepness of the potential peak. Therefore, eq.~(\ref{beyond-eikonal-K-f}) can be applied to an arbitrary effective potential and the resulting error must be small if the eikonal approximation is good for the particular form of the effective potential.

\begin{figure*}
\resizebox{\linewidth}{!}{\includegraphics{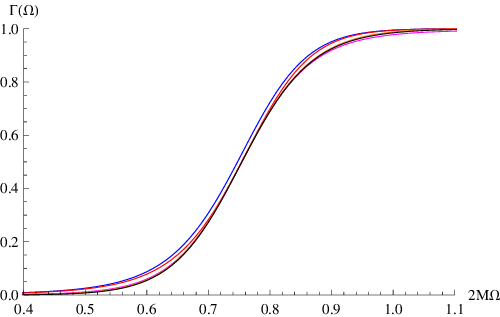}\includegraphics{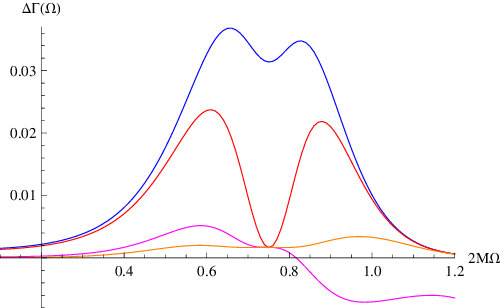}}
\caption{Transmission coefficients for the gravitational perturbations, $\ell=2$ ($2M\omega_0 = 0.747343 - 0.177925 \imo$, $2M\omega_1 = 0.693422 - 0.547830 \imo$). Left panel: the approximate values by the eikonal formula (blue), by the first-order formula (\ref{beyond-eikonal-K-1}) (red), by the second-order formula (\ref{beyond-eikonal-K-2}) (magenta), the amended formula (\ref{beyond-eikonal-K-f}) (orange), and the accurate value (black). Right panel: difference between the approximate and accurate values.}\label{fig:gravl2}
\end{figure*}

\begin{figure*}
\resizebox{\linewidth}{!}{\includegraphics{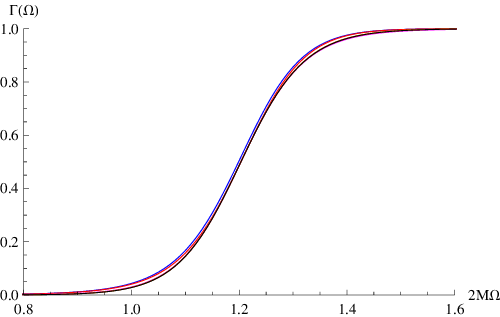}\includegraphics{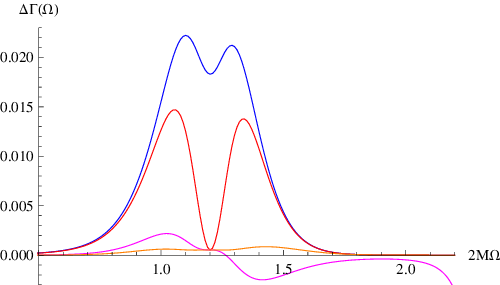}}
\caption{Transmission coefficients for the gravitational perturbations, $\ell=3$ ($2M\omega_0 = 1.198887 - 0.185406 \imo$, $2M\omega_1 = 1.165288 - 0.562596 \imo$). Left panel: the approximate values by the eikonal formula (blue), by the first-order formula (\ref{beyond-eikonal-K-1}) (red), by the second-order formula (\ref{beyond-eikonal-K-2}) (magenta), the amended formula (\ref{beyond-eikonal-K-f}) (orange), and the accurate value (black). Right panel: difference between the approximate and accurate values.}\label{fig:gravl3}
\end{figure*}

\section{Application to the Schwarzschild black hole}\label{sec:results}

The metric has the form \cite{Schwarzschild:1916uq},
\begin{equation}
ds^2=-f(r)dt^2+\frac{dr^2}{f(r)}+r^2(d\theta^2+\sin^2\theta d\phi^2),
\end{equation}
where
$$f(r)=1-\frac{2M}{r},$$
and $M$ is the black-hole mass.

The effective potential for axial types of gravitational perturbations has the form \cite{Regge:1957td},
\begin{equation}
V(r)=f(r)\left(\frac{\ell(\ell+1)}{r^2}-\frac{6M}{r^3}\right), \qquad \ell=2,3,4,\ldots.
\end{equation}

Since axial perturbations are isospectral to polar ones, we will focus exclusively on the first type of perturbations.

As we see from the figures \ref{fig:gravl2} and \ref{fig:gravl3}, the larger is the multipole number $\ell$, the more accurate is the correspondence. Already for $\ell=3$ gravitational perturbations of the Schwarzschild spacetime the approximate value of the grey-body factors obtained via the correspondence differs from the accurate ones only by less than $0.001$. It is essential that, when applying the correspondence to finding the grey-body factors, we use the accurate values of the quasinormal frequencies, $\omega_0$ and $\omega_1$,\footnote{The accurate values of the quasinormal frequencies of the Schwarzschild black hole are available from \url{http://qnms.way.to/}.} found by the Leaver method \cite{Leaver:1985ax}. To compare the grey-body factors obtained via the correspondence with the precise ones we used the numerical solution of the wave equation (\ref{wavelike-scatter}).

Thus, the correspondence can be used to estimate the grey-body factors, which are typically obtained by numerically solving the wave equation (\ref{wavelike-scatter}) and determining the reflection and transmission coefficients for each particular value of $\Omega$. For simpler black-hole geometries, it is possible to derive approximate analytic expressions for the grey-body factors by matching the near-horizon solution with the superposition of asymptotically ingoing and outgoing waves (see, e.g.,~\cite{Page:1976df}). The analytic formula (\ref{beyond-eikonal-K-f}) provides an approximation for grey-body factors based on the dominant quasinormal modes corresponding to the same type of perturbations, making it applicable to a wide range of black holes for which the WKB approximation can be used. These precise grey-body factors are shown in black color in figures \ref{fig:gravl2} and \ref{fig:gravl3}.

\section{Conclusions}\label{sec:conclusions}

We have demonstrated that the grey-body factors of spherically symmetric and asymptotically flat black holes in the regime of high multipole numbers $\ell$ are related to the values of the fundamental quasinormal mode via the remarkably simple equation \ref{transmission-eikonal}. This correspondence arises from the first-order WKB expansion and we further refined it by using higher-order WKB approximations. In the general case, the fundamental mode and the first overtone determine the grey-body factors, which can be accurately found through this correspondence, even at the lowest values of $\ell$ for gravitational perturbations.

The correspondence is limited to the class of effective potentials for which the WKB expansion is valid. Therefore, for effective potentials with an unusual form of the centrifugal term, such as those in Einstein-Gauss-Bonnet theory, the WKB formula does not work properly, as shown in \cite{Konoplya:2020bxa,Konoplya:2017wot}. The correspondence also applies to asymptotically de Sitter black holes, with the caveat that one should use the fundamental mode and first overtone of the de Sitter branch \cite{Konoplya:2022gjp} in such cases. It should also be applicable to higher-dimensional black holes without any modifications.

We note that the explicit form of the higher-order approximate formula should depend also on higher overtones. However, since we assume that the higher orders add only small corrections, we conclude that the dominant contribution to the grey-body factors comes from the dominant modes. Supposing that this is true in the general case and without approximation, unlike our approach which uses the WKB approximation and cannot reach the $\ell \lesssim n$ regime, it becomes clear why grey-body factors are much less sensitive than higher overtones to small deformations of the effective potential \cite{Rosato:2024arw}. The simple fact that grey-body factors do not depend on the highly sensitive higher overtones would provide this kind of stability.

Establishing a similar correspondence for a class of rotating black holes is a much more involved problem due to several factors. Firstly, to analyze the eikonal regime, it is necessary for the perturbation equations to allow for separation of variables, at least in the eikonal limit. Extending the correspondence beyond the eikonal regime would require separation of variables for all $\ell$. Additionally, the effective potential in this case is complex and depends on $\Omega$, necessitating substantial modifications to the above procedure. In a forthcoming paper, we plan to extend the correspondence to at least some class of axially symmetric black holes.

\Styleopt{}{\bigskip}
\bibliographystyle{unsrt}
\bibliography{Bibliography}
\end{document}